\documentclass[aps,prx,twocolumn,superscriptaddress]{revtex4-2}
\usepackage{tikz}
\usepackage{amsthm}
\usepackage{braket}
\usepackage{amsmath}
\usepackage{dashrule}
\usepackage{graphicx}
\usepackage{subfigure}
\usepackage[charter,cal=cmcal,sfscaled=false]{mathdesign}

\definecolor{iffsred}{cmyk}{0,0.9,0.8,0.4}
\definecolor{uestcblue}{cmyk}{1,0.8,0,0}
\usepackage{hyperref}
\hypersetup{
  pdftitle={Cost of Local Approximations of High-Dimensional Ground States Contextual Quantum Models},
  pdfauthor={Yang, Zhu, et al.},
  pdfstartview=Fit,
  pdfpagelayout=SinglePage,
  colorlinks,
  linkcolor=uestcblue,
  citecolor=uestcblue,
  urlcolor=iffsred}
\usepackage[figure]{hypcap}

\begin{document}
\title{Cost of locally approximating high-dimensional ground states of contextual quantum models}

\author{Kaiyan Yang}\thanks{These authors contributed equally}
\affiliation{Institute of Fundamental and Frontier Sciences, University of Electronic Science and Technology of China, Chengdu, 611731, China}
\affiliation{Key Laboratory of Quantum Physics and Photonic Quantum Information, Ministry of Education, University of Electronic Science and Technology of China, Chengdu, 611731, China}
\author{Yanzheng Zhu}\thanks{These authors contributed equally}
\affiliation{Institute of Fundamental and Frontier Sciences, University of Electronic Science and Technology of China, Chengdu, 611731, China}
\affiliation{Key Laboratory of Quantum Physics and Photonic Quantum Information, Ministry of Education, University of Electronic Science and Technology of China, Chengdu, 611731, China}
\author{Xiao Zeng}
\affiliation{Institute of Fundamental and Frontier Sciences, University of Electronic Science and Technology of China, Chengdu, 611731, China}
\affiliation{Key Laboratory of Quantum Physics and Photonic Quantum Information, Ministry of Education, University of Electronic Science and Technology of China, Chengdu, 611731, China}

\author{Zuoheng Zou}
\author{Man-Hong Yung}
\affiliation{Central Research Institute, Huawei Technologies, Shenzhen, 518129, China}

\author{Zizhu Wang}\email{zizhu@uestc.edu.cn}
\affiliation{Institute of Fundamental and Frontier Sciences, University of Electronic Science and Technology of China, Chengdu, 611731, China}
\affiliation{Key Laboratory of Quantum Physics and Photonic Quantum Information, Ministry of Education, University of Electronic Science and Technology of China, Chengdu, 611731, China}

\begin{abstract}{
Contextuality, one of the strongest forms of quantum correlations, delineates the boundary between the quantum world and the classical one. Recent advances show that some translation-invariant contextuality witnesses are maximally violated by ground states and local observables of infinite one-dimensional translation-invariant Hamiltonians. However, these models all have local Hilbert space dimension larger than two, making the study of their ground states behavior difficult on current qubit-based platforms. In this work, we focus on the cost of simulating their 3-site reduced density matrices using qubit-based parameterized quantum circuits. The local approximations are purified then encoded into permutation-symmetric qubit states. By developing a universal set of permutation-symmetry preserving qubit-based gates, we assess the accuracy of simulating the purified local ground states against fixed classical and quantum resources.  Results reveal that more contextual ground states with lower energy density are easier to simulate under identical resources. }
\end{abstract}

\maketitle

\section{Introduction}
Exploring non-classical correlations present in many-body quantum systems has been an impetus for quantum simulation since its inception. While entanglement has always been the primary focus of this endeavor, the study of a stronger form of quantum correlation—contextuality—has recently been gaining popularity. Contextuality, both the Kochen-Specker type~\cite{Budroni2022} and the Bell type~\cite{RevModPhys.86.419,Cabello2021KSBell}, has been shown to be the source of quantum advantages in specific tasks in quantum computing~\cite{Howard2014-Q-adavantage-contextuality-magic, Bravyi2018-Q-advantage-shallow-circuit-nonlocality} and quantum machine learning~\cite{GaoXun2023-Q-advantage-contextuality-nerual-learning}. 

In an earlier work by some of the authors~\cite{Yang2022}, we found several infinite one-dimensional translation-invariant (TI) quantum Hamiltonians with nearest- and next-to-nearest-neighbor interactions having the lowest ground state energy density permissible in quantum physics. This lower bound is numerically certified by a variant of the NPA-hierarchy~\cite{Miguel2007,Miguel2008}, and the quantum Hamiltonian achieving the bound has been found by optimizing the local observables of contextuality witnesses derived from Bell inequalities and computing their ground states with uniform matrix product state (uMPS) algorithms~\cite{Haegeman2011-TDVP,ZaunerStauber2018-VUMPS-short,Vanderstraeten2019-VUMPS-long,Cirac2021}. These quantum Hamiltonians, which are essentially many-body quantum Bell inequalities~\cite{tsirelson1980bound}, demarcate quantum and post-quantum theories, providing a conceptual way of experimentally refuting the universal validity of quantum physics~\cite{Yang2022}. In addition, they can be used in self-testing protocols~\cite{supicSelftestingQuantumSystems2020} for many-body quantum systems, which has recently been done for entangled states in a network scenario~\cite{Supic2023NetworkSelfTesting}. Self-testing can be seen as a method to separate classical control from quantum behavior~\cite{Reichardt2013SelfTesting}, leading to the central question of this work: How hard is it to generate local approximations of ground states of quantum Hamiltonians that reach the lower bound?

In order to answer this question, we use variational quantum algorithms (VQAs)~\cite{Cerezo2021VQA} to simulate the purification of local reduced density matrices of ground states of quantum Hamiltonians exhibiting various degrees of contextuality. However, since these Hamiltonians all have local Hilbert space dimension 3 but our variational simulator is based on qubits, we encode a qutrit state into a combination of permutation-symmetric qubits. Exploiting permutation symmetry in many-body quantum systems allows us to characterize various forms of quantum correlations in many scenarios~\cite{Tura2014,Tura2015,PhysRevE.78.021106,PhysRevLett.106.180502,PhysRevA.83.042332,PhysRevLett.108.210407,aulbach2010maximally}. We also develop a qubit-based ansatz inspired by recent ansatzes based on tensor network states~\cite{Ran2020,Rudolph2023}. Compared to earlier works using qubit-based variational algorithms to simulate higher-dimensional quantum systems~\cite{Gard2020-symmetric-preserve-Q-circuit,Sawaya2020,Lyu2023,Meyer2023,Han2024}, our ansatz preserves the permutation symmetry of the encoded states and allows arbitrary 2-qutrit gates to be implemented. 

Using these techniques, we investigate the cost of locally simulating qutrit ground states of contextual quantum Hamiltonians from Yang et al.~\cite{Yang2022} via qubit-based parameterized quantum circuits (PQCs). The cost is quantified based on the consumption of classical resources and quantum resources, with its evaluation metric defined by comparing the negative logarithmic fidelity per site (NLF-per-site) of the PQC-generated state and the purified local ground state. Concretely, the classical resources are quantified using three indicators: (i) the number of classical variational parameters in the circuit, (ii) the number of iterations in classical optimization component of the VQA, and (iii) the magnitude of bond dimension representing the target state. Additionally, the quantum resource is quantified by the number of quantum gates employed. A lower NLF-per-site signifies the two states are closer, making the simulation better. Then, preparing two different purified states under the same circuit, classical and quantum resources, the state with a lower NLF-per-site can be simulated easier than the state with a higher  NLF-per-site.

Two observations stand out from our results. First, by comparing the preparation costs of quantum states with varying degrees of contextuality, we observe that a greater violation of the contextuality witness corresponds to a lower NLF-per-site for preparing the respective ground state, suggesting that ground states with stronger contextuality are easier to simulate. Therefore, the ground state energy density can serve as a reliable indicator of the cost of approximating the local ground states with varying degrees of contextuality. Second, when comparing the preparation costs of noncontextual states with those of all contextual quantum states, we find that all contextual ground states exhibit a higher NLF-per-site than the noncontextual ground state whose energy density reaches the classical bound. This indicates that preparing contextual ground states is generally more expensive than preparing noncontextual ones, which are often separable states.

\section{Methods} 
\subsection{Quantum Hamiltonians and reduced ground states} \label{sec:purification}
The Hamiltonians we consider in this work are infinite one-dimensional translation-invariant with nearest- and next-to-nearest neighbor interactions with two dichotomic observables $\sigma_x,\sigma_y$ per site: 
\begin{equation}
    \begin{split}
        H & = \sum_{i=1}^{\infty} J_x{\sigma_x^i} + J_y{\sigma_y^i} + J_{xx}^{AB}{\sigma_x^i \sigma_y^{i+1}} + J_{xy}^{AB}{\sigma_x^i \sigma_y^{i+1}} \\ 
        & 
        +  J_{yx}^{AB}{\sigma_y^i \sigma_x^{i+1}} + J_{yy}^{AB}{\sigma_y^i \sigma_y^{i+1}} + J_{xx}^{AC}{\sigma_x^i \sigma_x^{i+2}} \\
        &
        + J_{xy}^{AC}{\sigma_x^i \sigma_y^{i+2}} + J_{yx}^{AC}{\sigma_y^i \sigma_x^{i+2}} +
        J_{yy}^{AC}{\sigma_y^i \sigma_y^{i+2}}.
    \end{split}
    \label{Hamiltonian322}
\end{equation}

This class of Hamiltonians is called the 322-type in Yang et al.~\cite{Yang2022}. The coupling constants ${\cal{J}} \equiv \{J_x, J_y, J_{xx}^{AB}, J_{xy}^{AB}, J_{yx}^{AB}, J_{yy}^{AB}, J_{xx}^{AC},  J_{xy}^{AC}, J_{yx}^{AC}, J_{yy}^{AC}\}$ come from contextuality witnesses of the form

\begin{equation}
    \begin{split}
       & J_x{\langle O_x^1 \rangle} + J_y{\langle O_y^1 \rangle} + J_{xx}^{AB}{\langle O_x^1 O_x^2 \rangle} + J_{xy}^{AB}{\langle O_x^1 O_y^2 \rangle} 
         \\ & + J_{yx}^{AB}{\langle O_y^1 O_x^2 \rangle} + J_{yy}^{AB}{ \langle O_y^1 O_y^2 \rangle} + J_{xx}^{AC}{\langle O_x^1 O_x^3 \rangle} 
        \\ & + J_{xy}^{AC}{\langle O_x^1 O_y^3 \rangle} + J_{yx}^{AC}{\langle O_y^1 O_x^3 \rangle} +
        J_{yy}^{AC}{\langle O_y^1 O_y^3 \rangle} \geq {\cal{L}}.
    \end{split}
    \label{322 cw form}
\end{equation}
where $\cal{L}$ is the classical bound, the violation of which indicates the presence of contextuality. The Hamiltonians in Eq.(\ref{Hamiltonian322}) are constructed from these witnesses by replacing $O_x,O_y$ with local observables $\sigma_x,\sigma_y$, which can be any Hermitian operator with spectrum contained in $\{-1, 1\}$. In one dimension, these contextuality witnesses can be completely characterized by convex polytopes, when the maximum interaction distance (3 in our case because we include next-to-nearest neighbor interactions), the number of local observables (2 in our case) and the number of outcomes per observable (also 2 in our case) is fixed~\cite{Yang2022,Wang2017}.

Given the coefficients ${\cal{J}}$, if there exists a set of values for $\{\sigma_x,\sigma_y\}$ such that the ground state energy density of Hamiltonian \eqref{Hamiltonian322} is smaller than $\cal{L}$, it indicates the contextuality witness \eqref{322 cw form} is violated by the ground state and local observables $\{\sigma_x,\sigma_y\}$. Results in Yang et al.~\cite{Yang2022} reveal that among the 2102 inequivalent contextuality witnesses, five can be maximally violated when the local Hilbert space dimension being $3$ (see Table~\ref{table:5 CW}).

\begin{table*}
    \caption{Five maximally violated contextuality witnesses. In Yang et al.~\cite{Yang2022}, five contextuality witnesses are maximally violated by the ground states and local observables of five 322-type 1D infinite translation-invariant Hamiltonians when the local Hilbert space dimension being 3. Here, ${\mathcal{L}}$ is classical bound, ${\cal{J}} \equiv \{J_x, J_y, J_{xx}^{AB}, J_{xy}^{AB}, J_{yx}^{AB}, J_{yy}^{AB}, J_{xx}^{AC},  J_{xy}^{AC}, J_{yx}^{AC}, J_{yy}^{AC}\}$ are the coefficients of contextuality witness which are also couplings of the Hamiltonian \eqref{Hamiltonian322}, and ${\cal{Q}}_{\rm{Limit}}$ is the maximal quantum violation obtained when the local physical dimension is $3$. }
    \renewcommand\arraystretch{1.4}
    \setlength{\tabcolsep}{2ex}
    \begin{ruledtabular}
    \begin{tabular}{ccccccccccccccc}
        No. & ${\cal{L}}$ & $J_x$ &$J_y$ & $J_{xx}^{AB}$ & $J_{xy}^{AB}$ & $J_{yx}^{AB}$ & $J_{yy}^{AB}$ & $J_{xx}^{AC}$ & $J_{xy}^{AC}$
        & $J_{yx}^{AC}$ & $J_{yy}^{AC}$ & ${\cal{Q}}_{\rm{Limit}}$ \\ \hline
        1 & -6 & -6 & 0 & 2 & 3 & 3 & -2 & 3 & -1 & -1 & 1 & -6.32747 \\
        2 & -6 & -4 & 2 & 2 & 2 & 2 & -4 & 1 & -1 & -1 & 3 & -6.33712 \\
        3 & -3 & -3 & 1 & 1 & 1 & 1 & -1 & 1 & 0 & -1 & 1 & -3.20711 \\
        4 & -4 & -2 & -2 & -2 & 1 & -1 & -2 & 1 & 0 & 2 & 1 & -4.14623	\\
        5 & -8 & -11 & 1 & 5 & 2 & 2 & -1 & 4 & -1 & -2 & 1 & -8.12123
    \end{tabular}
    \end{ruledtabular}
    \label{table:5 CW}
\end{table*}

The ground state of Hamiltonian~\eqref{Hamiltonian322} is represented by a uMPS. A uMPS $|\phi(A)\rangle$ with the bond dimension $D$ and the physical dimension $d$ defined on an infinite 1D TI chain is parameterized by a set of $D\times D$ matrices $A^s (s=1,2,\dots,d)$. The overall TI variational ground state then can be written as 
\begin{equation}
    |\phi(A) \rangle = \sum_{\mathbf{s}} (\cdots A^{s_{i-1}} A^{s_i} A^{s_{i+1}} \cdots)|\mathbf{s} \rangle,
    \label{eq:umps}
\end{equation}
and represented diagrammatically as
\begin{equation*}\label{c4-umps}
    \begin{split}
        \begin{tikzpicture}
            \node (begin){$|\phi(A)\rangle = \cdots$};
            \node(m1)[draw, rounded corners, right of = begin, xshift = 0.8cm]{$A$};
            \node(m2)[draw, rounded corners, right of = m1, xshift = 1mm]{$A$};
            \node(m3)[draw, rounded corners, right of = m2, xshift = 1mm]{$A$};
            \node(m4)[draw, rounded corners, right of = m3, xshift = 1mm]{$A$};
            \node(m5)[draw, rounded corners, right of = m4, xshift = 1mm]{$A$};
            \node(end)[right of = m5, xshift = 0.5mm]{$\cdots$};
            \node(b1)[below of = m1, yshift = 2mm]{};
            \node(b2)[below of = m2, yshift = 2mm]{};
            \node(b3)[below of = m3, yshift = 2mm]{};
            \node(b4)[below of = m4, yshift = 2mm]{};
            \node(b5)[below of = m5, yshift = 2mm]{};
            \draw (m1)--(begin.east);
            \draw (m1)--(m2)--(m3)--(m4)--(m5);
            \draw (m5)--(end.west);
            \draw (m1)--(b1.north);
            \draw (m2)--(b2.north);
            \draw (m3)--(b3.north);
            \draw (m4)--(b4.north);
            \draw (m5)--(b5.north);
        \end{tikzpicture}
    \end{split}
\end{equation*} 

To explore the cost of preparing the ground state via quantum circuits, we simulate its local reduced state. Since the 322-type local Hamiltonian operates on three adjacent sites, we need at least 3-site reduced state to calculate its ground state energy density. We focus on the minimal case and the 3-site reduced state of the ground state $|\phi(A)\rangle$ is
\begin{equation}\label{3-site RDM}
    \begin{split}
        \begin{tikzpicture} 
            \node(m1) at(0.7,0) {$\rho_3 = $};
            \node(l)[draw, circle, scale=0.8] at(1.4,0){$l$};
            \node(m3)[draw, rounded corners]at(2.3,1){$A$};
            \node(m4)[draw, rounded corners]at(3.5,1){$A$};
            \node(m5)[draw, rounded corners]at(4.7,1){$A$};
            \node(n3)[draw, rounded corners]at(2.3,-1){$\bar{A}$};
            \node(n4)[draw, rounded corners]at(3.5,-1){$\bar{A}$};
            \node(n5)[draw, rounded corners]at(4.7,-1){$\bar{A}$};				
            \node(r)[draw, circle, scale=0.8] at(5.6,0){$r$};
            \draw [-](l) to [out=90,in=180] (m3);
            \draw [-](l) to [out=270,in=180] (n3);
            \draw [-](r) to [out=90,in=0] (m5);
            \draw [-](r) to [out=270,in=0] (n5);
            \draw (m3.south)--(2.3,0.3);
            \draw (m4.south)--(3.5,0.3);
            \draw (m5.south)--(4.7,0.3);
            \draw (n3.north)--(2.3,-0.3);
            \draw (n4.north)--(3.5,-0.3);
            \draw (n5.north)--(4.7,-0.3);
            \draw (m3) -- (m4);
            \draw (m4) -- (m5);
            \draw (n3) -- (n4);		
            \draw (n4) -- (n5);	
        \end{tikzpicture}
    \end{split}
\end{equation}
Here, $l$ and $r$ are the left and right fixed points of the transfer matrix $T = \sum_s {\bar{A}}^s \otimes A^s $ with non-degenerate leading eigenvalue 1. Diagrammatically, $l$ and $r$ satisfy
\begin{equation*}\label{LR fixed point}
    \begin{split}
        \begin{tikzpicture} 
            \node(l)[draw, circle, scale=0.8] at(0.5,0){$l$};
            \node(m)[draw, rounded corners]at(1.4,1){$A$}; 
            \node(n)[draw, rounded corners]at(1.4,-1){$\bar{A}$};    
            \draw [-](l) to [out=90,in=180] (m);
            \draw [-](l) to [out=270,in=180] (n);      
            \draw (m.south)--(n.north);              
            \draw (m) -- (2,1);           
            \draw (n) -- (2,-1);		
            \node(eq) at(2.3,0) {$ = $};
            \node(L)[draw, circle, scale=0.8] at(3,0){$l$};
            \draw [-] (L) to [out=90,in=180] (3.7,1);
            \draw [-] (L) to [out=270,in=180] (3.7,-1);       
            \node(mm)[draw, rounded corners]at(4.5+1,1){$A$}; 
            \node(nn)[draw, rounded corners]at(4.5+1,-1){$\bar{A}$};
            \draw (mm.south)--(nn.north);
            \draw (mm.west)--(4+0.9,1);
            \draw (nn.west)--(4+0.9,-1);
            \node(R)[draw, circle, scale=0.8] at(5.5+0.9,0){$r$};
            \draw [-](R) to [out=90,in=0] (mm);
            \draw [-](R) to [out=270,in=0] (nn);
            \node(EQ) at(7.2,0) {$ = $};
            \node(RR)[draw, circle, scale=0.8] at(7+1.2,0){$r$};
            \draw [-](RR) to [out=90,in=0] (7.5,1);
            \draw [-](RR) to [out=270,in=0] (7.5,-1);
        \end{tikzpicture}
    \end{split}
\end{equation*}
Besides, $l$ and $r$ satisfy the normalization condition $\mathrm{Tr}(lr)=1$, or diagrammatically,
\begin{equation*}\label{LR normalization}
    \begin{split}
        \begin{tikzpicture} 
            \node(l)[draw, circle, scale=0.8] at(0.5,0){$l$};
            \node(r)[draw, circle, scale=0.8] at(0.5+1.3,0){$r$};
            \draw [-](l) to [out=90,in=90, distance=1cm] (r);
            \draw [-](r) to [out=270,in=270,distance=1cm] (l);
        \end{tikzpicture}
    \end{split}=1
\end{equation*}

Given that $\rho_3$ is generally mixed, to prepare it via PQC, we use ancilla qutrits to purify $\rho_3$ into a pure state $\ket{\psi}$. First, we perform Cholesky decomposition on $l$ and $r$: 
$$l=L^{\dagger}L, \quad r=RR^{\dagger}.$$ 
$\rho_3$ is then represented by 
\begin{equation*}\label{3-site RDM cholesky}
    \begin{split}
        \begin{tikzpicture} 
            \node(m1) at(0.7,0) {$\rho_3 = $};
            \node(L)[draw, circle, scale=0.8] at(1.4,0.4){$L$};
            \node(Ldagger)[draw, circle, scale=0.7] at(1.4,-0.4){$\bar{L}$};
            \node(m3)[draw, rounded corners]at(2.3,1){$A$};
            \node(m4)[draw, rounded corners]at(3.5,1){$A$};
            \node(m5)[draw, rounded corners]at(4.7,1){$A$};
            \draw [-](L) to [out=90,in=180] (m3);
            \draw [-](Ldagger) to [out=270,in=180] (n3);
            \draw (Ldagger) -- (L);
            \node(n3)[draw, rounded corners]at(2.3,-1){$\bar{A}$};
            \node(n4)[draw, rounded corners]at(3.5,-1){$\bar{A}$};
            \node(n5)[draw, rounded corners]at(4.7,-1){$\bar{A}$};				
            \node(R)[draw, circle, scale=0.8] at(5.6,0.4){$R$};
            \node(Rdagger)[draw, circle, scale=0.7] at(5.6,-0.4){$\bar{R}$};
            \draw [-](R) to [out=90,in=0] (m5);
            \draw [-](Rdagger) to [out=270,in=0] (n5);
            \draw (Rdagger) -- (R);
            \draw (m3.south)--(2.3,0.3);
            \draw (m4.south)--(3.5,0.3);
            \draw (m5.south)--(4.7,0.3);
            \draw (n3.north)--(2.3,-0.3);
            \draw (n4.north)--(3.5,-0.3);
            \draw (n5.north)--(4.7,-0.3);
            \draw (m3) -- (m4);
            \draw (m4) -- (m5);
            \draw (n3) -- (n4);		
            \draw (n4) -- (n5);	
        \end{tikzpicture}
    \end{split}
\end{equation*}
Then, we split $\rho_3$ into top and bottom parts. The top part is the target state $|\psi\rangle$
\begin{equation*}\label{RDM tmp tensor}
    \begin{split}
        \begin{tikzpicture} 
            \node(m1) at(0.2,-0.3) {$|\psi\rangle = $};
            \node(L)[draw, circle, scale=0.8]at(1.1,0){$L$};
            \node(L1-dimension) [scale=0.8] at (1.3,-0.7){$D$};
            \node(L2-dimension) [scale=0.8] at (1.7,0.25){$D$};
            \node(m3)[draw, rounded corners]at(2.3,0){$A$};
            \node(m4)[draw, rounded corners]at(3.5,0){$A$};
            \node(m5)[draw, rounded corners]at(4.7,0){$A$};
            \node(R)[draw, circle, scale=0.8]at(5.9,0){$R$};
            \node(R1-dimension) [scale=0.8] at (6.1,-0.7){$D$};
            \node(R2-dimension) [scale=0.8] at (5.3,0.25){$D$};
            \draw (L) -- (m3);                  			       
            \draw (R) -- (m5);
            \draw (L.south) -- (1.1,-0.7);
            \draw (m3.south)--(2.3,-0.7);
            \draw (m4.south)--(3.5,-0.7);
            \draw (m5.south)--(4.7,-0.7);
            \draw (R.south) -- (5.9,-0.7);
            \draw (m3) -- (m4);
            \draw (m4) -- (m5);
        \end{tikzpicture}
    \end{split}
\end{equation*} 
However, the physical dimension of $L$ and $R$ is not $d$, so we express $L$ and $R$ using $k=\lceil \text{log}_dD\rceil$ tensors with physical dimension $d$ (assuming $D>d$). If $\text{log}_dD=k$ is an integer, we repeatedly apply singular value decompositions (SVDs) to $L$ and $R$
\begin{equation*}\label{L-SVD}
    \begin{split}
        \begin{tikzpicture} 
            \node(L)[draw, rounded corners] at(0,0) {$L$};
            \node() [scale=0.8] at(0.2,-0.7) {$d^k$};
            \draw (L) -- (0.8,0);
            \draw (L.south) -- (0,-0.7);
            \node at(1.3,0) {$\longrightarrow$};
            \node[scale=0.8] at(1.3,0.3) {SVD};
            \node [scale=0.8] at(0.5,0.3) {$D$};
            \node(l2)[draw, rounded corners]at (2.2,0){$L_2$};
            \node [scale=0.8] at (3.6,-0.7){$d$};
            \node [scale=0.8] at (2.8,0.25){$d$};
            \node [scale=0.8] at (4,0.25){$D$};
            \node [scale=0.8] at (2.6,-0.7){$d^{k-1}$};
            \node(l1)[draw, rounded corners]at(3.4,0){$L_1$};
            \draw (l2) -- (l1);
            \draw (l1) -- (3.4+0.8,0);
            \draw (l2.south) -- (2.2,-0.7);
            \draw (l1.south) -- (3.4,-0.7);
            \node at(4.5,0) {$\cdots$};
            \node at(5.1,0) {$\longrightarrow$};
            \node[scale=0.8] at(5.1,0.3) {SVD};
            \node(l2)[draw, rounded corners]at (2.2+3.7,0){$L_k$};
            \draw (l2) -- (2.8+3.7,0);
            \node [scale=0.8] at (3.6+4.3,-0.7){$d$};
            \node [scale=0.8] at (2.8+4,0){$\cdots$};
            \node [scale=0.8] at (2.4+3.7,-0.7){$d$};
            \node(l1)[draw, rounded corners]at(3.4+4.3,0){$L_1$};
            \draw (l1) -- (2.8+4.3,0);
            \draw (l1) -- (3.4+0.6+4.3,0);
            \draw (l2.south) -- (2.2+3.7,-0.7);
            \draw (l1.south) -- (3.4+4.3,-0.7);        
        \end{tikzpicture}
    \end{split}
\end{equation*}
Similarly, we repeatedly do SVDs on $R$ and get tensors $\{R_1,\cdots,R_k\}$. Replacing $L$ and $R$ with $\{L_1,\cdots,L_k\}$ and $\{R_1,\cdots,R_k\}$ respectively, we finally get the purified state. If $\text{log}_dD$ is not an integer, we enlarge the physical dimension of $L$ to $d^k$ by adding zeros to $L$, and then repeatedly do SVDs as before.
Note that, the mixed state $\rho_3$ can be reconstructed by tracing out the ancilla qutrits from both the leftmost $k$-site and the rightmost $k$-site.

\subsection{Simulating qutrits} \label{sec:method}
\begin{figure*}
    \centering
    \includegraphics[width=\textwidth]{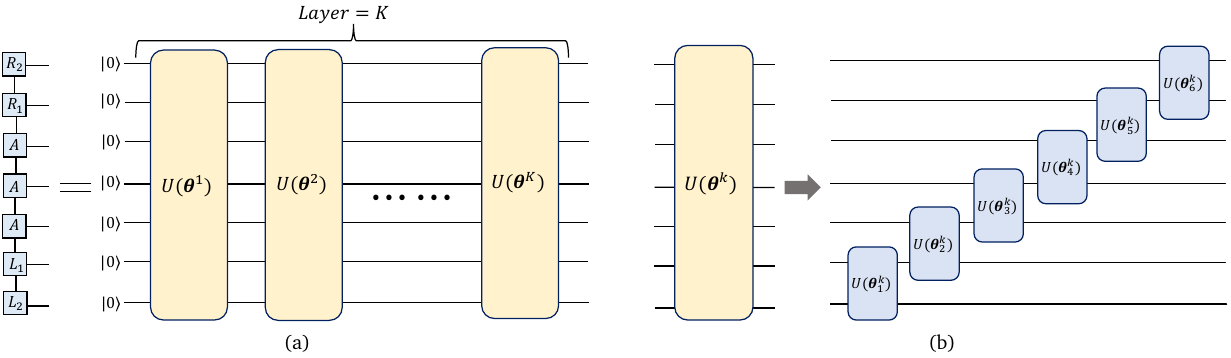}
    \caption{Quantum circuit with $K$ layers for a 7-site qutrit matrix product state. (a) An illustration of a $K$-layer quantum circuit ($\textit{Layer}=k$). Each block unitary $U(\boldsymbol{\theta}^k)$ ($k \in \{1,2,\dots, K\}$) represents one layer of the circuit. (b) Structure of block unitary $U({\boldsymbol{\theta}}^k)$ in circuit (a). Each $U({\boldsymbol{\theta}}^k)$ consists of 6 parameterized two-qutrit unitaries $\{U({\boldsymbol{\theta}}_i^k)|i=1,2,\dots,6\}$, where $\boldsymbol{\theta}_i^k$ is a set of parameters and ${\boldsymbol{\theta}}^k = \{\boldsymbol{\theta}_i^k|i=1,2,\dots,6\}$ is collection of parameter sets.}
    \label{fig:7-site-quantum-circuit}
\end{figure*}

Even though the purification procedure above works for any local Hilbert space dimension, the Hamiltonians maximally violating the five witnesses in Table~\ref{table:5 CW} all have local dimension 3, which will be our focus for the rest of this work. After obtaining the purified qutrit MPS via the purification procedure, we need to variationally prepare them using PQCs. A good ansatz allows the simulation to be done efficiently and accurately. The ansatz we use consists of stacked linear layers of two-qutrit parameterized unitaries, similar to the one used for qubits in Ran~\cite{Ran2020} and Rudolph et al.~\cite{Rudolph2023}. We purify the 3-site local reduced state of bond dimension $D\in \{5,6,7,8,9\}$ to a 7-site pure state $|\psi(A,L_1,L_2,R_1,R_2)\rangle$, and the quantum circuit with $K$ layers is given by Figure~\ref{fig:7-site-quantum-circuit},
where the $k$-layer consists of a stack of 6 parameterized two-qutrit unitaries $\{U({\boldsymbol{\theta}}_i^k)\}$ for $i\in \{1,2,\dots,6\}$ and $k\in\{1,2,\dots,K\}$. The ansatz above can approximate $|\psi\rangle$ up to very high accuracy, with proper optimization techniques~\cite{Rudolph2023}. It is also a typical case of the finite local-depth circuit (FLDC) described in Zhang et al. \cite{Zhang2024absence-BP-FLDC}, which is shown to be free from the barren plateau problem. 

However, in each layer of Figure~\ref{fig:7-site-quantum-circuit}, $\{U({\boldsymbol{\theta}_i^k})\}$ are qutrit unitaries. In order to simulate the purified local ground states on qubit-based platforms, we encode each qutrit state into a combination of permutation-symmetric qubit states and develop a universal set of permutation-symmetry preserving qubits gates to approximate any 2-qutrit gate. Existing encoding methods which allow the simulation of high-dimensional quantum systems on qubit-based VQAs typically use standard binary, Gray code, or unary encoding~\cite{Sawaya2020}. They do not guarantee that the encoded states stay in the encoded subspace through gate operations. Our gate operations preserve the symmetry of encoded qubits. Similar encoding methods based on symmetry has been applied to quantum chemistry simulation~\cite{Gard2020-symmetric-preserve-Q-circuit}, variational quantum spin eigensolver~\cite{Lyu2023}, variational quantum machine learning~\cite{Meyer2023}, and multilevel variational spectroscopy simulator~\cite{Han2024}. However, as noted below, our method differs from these one substantially.

The Hilbert space of a $d$-dimensional ($d\ge3$) qudit is isomorphic to the symmetric subspace $\mathcal{H}_s$ of the $n$-qubit system, where $n=d-1$ and $\mathcal{H}_s$ is spanned by the Dicke basis $\{\ket{S_m}:0\leq m\leq n\}$ \cite{Dicke1954}
\begin{equation}
    \ket{S_m}={\binom{n}{m}}^{-1/2}\sum_{\text{perm}}\ket{1}^{\otimes m}\ket{0}^{\otimes(n-m)}.
    \label{eq-Dicke states}
\end{equation}
$\ket{S_m}$ is the coherent superposition of all permutations of the computational basis states with $m$ qubits being $\ket{1}$ and $n-m$ qubits being $\ket{0}$. The Hilbert space for two qubits $\mathcal{H}_4$ can be decomposed into the direct sum of the three-dimensional symmetric subspace $\mathcal{H}_s$ and the one-dimensional anti-symmetric subspace $\mathcal{H}_a$ spanned by the anti-symmetric basis $\ket{S_a}=(\ket{01}-\ket{10})/{\sqrt{2}}$, i.e. $\mathcal{H}_4=\mathcal{H}_s\bigoplus\mathcal{H}_a$.

Denote the qutrit Hilbert space by $\mathcal{H}_3$ with basis $\{\ket{0},\ket{1},\ket{2}\}$. For a qutrit state $\ket{\psi}=\psi_0\ket{0}+\psi_1\ket{1}+\psi_2\ket{2}$, we define the symmetric encoded state as $\ket{\psi}_\text{enc} =\psi_0\ket{S_0}+\psi_1\ket{S_1}+\psi_2\ket{S_2}$, where $\ket{S_0}=\ket{00},\,\ket{S_1}=\frac{1}{\sqrt{2}}(\ket{01}+\ket{10}),\, \ket{S_2}=\ket{11}$.

Qubit gates may map a symmetric state $\ket{\psi}_\text{enc}$ out of $\mathcal{H}_s$. To make sure we always operate in $\mathcal{H}_s$ thus do not increase the total Hilbert space dimension of the VQA, we need gates that are symmetry-preserving. Even though Gard et al.~\cite{Gard2020-symmetric-preserve-Q-circuit} and Lyu et al.~\cite{Lyu2023} both incorporate symmetries into circuit design, the space they considered is the linear span of all permutations of the state $\ket{1}^{\otimes m}\ket{0}^{\otimes(n-m)}$ with a certain $m$. Moreover, the circuits they used are designed to preserve different kinds of symmetries (like total spin and particle number), which helps to improve accuracy and reduce computational resources. Our circuit design contains a universal gate set for simulating any qutrit gates using qubit gates, and the gate set is symmetry-preserving.

Let $U\in U(3)$ be a qutrit gate. Then, $U$ can be written as 
\begin{equation}
    U=\sum_{i,j=0}^{2}u_{ij}\ket{i}\bra{j}.
    \label{eq-qutrit-gate-qutrit-basis}
\end{equation}
Applying the symmetric state encoding on the computational basis states $\ket{i}$ and $\bra{j},\, i,j\in\{0,1,2\}$, $U$ can be encoded as $\tilde{U}$ in terms of Dicke basis $\{\ket{S_0}, \ket{S_1}, \ket{S_2}\}$:
\begin{equation}
    \tilde{U}=\sum_{i,j=0}^{2}u_{ij}\ket{S_i}\bra{S_j}.
    \label{eq-qutrit-gate-Dicke-basis}
\end{equation}

However, $\tilde{U}$ is not a unitary in $U(4)$ because the anti-symmetric component is missing. By adding a projector $P = \ket{S_a}\bra{S_a} $, where $\ket{S_a} = (\ket{01} - \ket{10})/{\sqrt{2}}$ is the anti-symmetric basis of $\mathcal{H}_a$, $U$ can be encoded as a two-qubit unitary $U_\text{enc}$:
\begin{equation}
    U_\text{enc}=\tilde{U}+P.
    \label{eq:Uenc}
\end{equation}
It can be checked that $U_\text{enc}$ is both unitary and symmetry-preserving.

The development of high-dimensional quantum computing lead to several proposals for high-dimensional universal gate sets~\cite{Gottesman1999QuditFTC,Muthukrishnan2000,Brennen2005,Bullock2005,Luo2014,Ringbauer2022-universal-qudit-Q-processor}. Di et al.~\cite{Di2013, Di2015} proposed an efficient method for synthesizing qudit unitaries using quantum Shannon decomposition~\cite{Shende2006}. In this method, high-dimensional gates are decomposed recursively, until each block only acts on a 2-dimensional subspace (i.e. a qubit). To transform the resulting blocks into an simpler form, Gaussian elimination is used to zero out off-diagonal entries. The end result of the procedure is a product of ``simple'' gates, each of which only acts on a 2-dimensional subspace of the high-dimensional unitary. In our case, for an arbitrary single-qutrit unitary $U\in U(3)$, this method decomposes it into a product of two-level qutrit unitary gates:
\begin{equation}
    U = V^{01}V^{02}V^{12},
    \label{decompose 1-qutrit unitary}
\end{equation}
where $V^{ij} \in U(3)$, $(i,j) \in \{(0,1),(0,2),(1,2)\}$ are two-level qutrit unitary gates with the general form
\begin{equation}
    V^{01}\!=\begin{bmatrix}
        {\theta}_1 & {\theta}_2 & 0 \\ {\theta}_3 & {\theta}_4 & 0 \\ 0 & 0 & 1
    \end{bmatrix},\;
    V^{02}\!=\begin{bmatrix}
        {\theta}_1 & 0 & {\theta}_2 \\ 0 & 1 & 0 \\ {\theta}_3 & 0 & {\theta}_4
    \end{bmatrix},\;
    V^{12}\!=\begin{bmatrix}
        1 & 0 & 0 \\ 0 & {\theta}_1 & {\theta}_2 \\ 0 & {\theta}_3 & {\theta}_4
    \end{bmatrix}.
\end{equation}
Here, ${\theta}_1,{\theta}_2,{\theta}_3$ and ${\theta}_4$ are the entries of an qubit unitary $V({\boldsymbol \theta}) \in U(2)$: 
\begin{equation}
    V({\boldsymbol \theta})=\begin{bmatrix}
        {\theta}_1 & {\theta}_2 \\ {\theta}_3 & {\theta}_4
    \end{bmatrix}.
    \label{qubit parameterized V}
\end{equation}

\begin{figure}
    \centering
    \includegraphics[width=\linewidth]{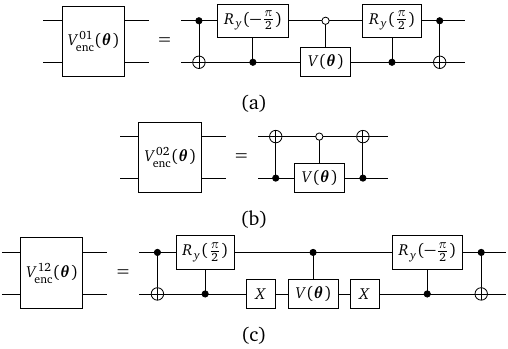}
    \caption{The qubit-based circuit for synthesizing two-level two-qubit symmetry-preserving unitaries. (a) Synthesis of $V^{01}_\text{enc}({\boldsymbol\theta})$. (b) Synthesis of $V^{02}_\text{enc}({\boldsymbol\theta})$. (c) Synthesis of $V^{12}_\text{enc}({\boldsymbol\theta})$. Here, $\bullet$ denotes the control state is $\ket{1}$, and $\circ$ denotes the control state is $\ket{0}$. $R_y$ represents the rotation-y gate, $X$ denotes Pauli $X$ gate, and $V(\boldsymbol \theta) \in U(2)$ has the general form of \eqref{qubit parameterized V}.}
    \label{fig-1-qutrit-encoded-systhesis}
\end{figure}

By implementing the gate encoding method on $V^{ij}({\boldsymbol \theta})$, we obtain the two-qubit symmetry-preserving unitary $V^{ij}_\text{enc}({\boldsymbol \theta})$. The circuit for synthesizing $V^{01}_\text{enc}({\boldsymbol \theta}), V^{02}_\text{enc}({\boldsymbol \theta})$ and $V^{12}_\text{enc}({\boldsymbol \theta})$ are shown in Figure~\ref{fig-1-qutrit-encoded-systhesis}.

\begin{figure*}
    \centering
    \includegraphics[width=\textwidth]{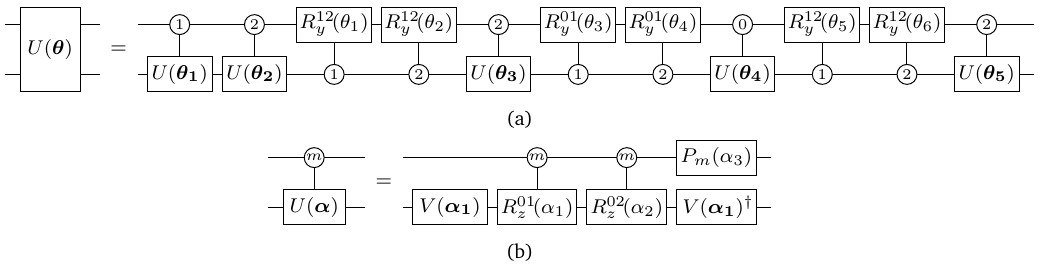}
    \caption{The qutrit-based circuit for synthesizing an arbitrary two-qutrit unitary $U \in U(9)$. (a) Synthesis of $U \in U(9)$ consisting of qutrit-controlled unitary gates and qutrit two-level controlled rotation gates. Here, \text{\scriptsize$\bigcirc$} denotes the control, with the number ($0,1$ or $2$) inside representing the state it is in. Note that each qutrit-controlled unitary gate can be further decomposed into single-qutrit gates and qutrit two-level controlled rotation gates, with the decomposition shown in (b). (b) Synthesis of qutrit controlled unitary gates in (a). Here, \text{\scriptsize$\bigcirc$} denotes the control in state $\ket{m}$ with $m=0,1,2$. $R_z^{01}$ and $R_z^{02}$ are qutrit two-level rotation gates, $P_m=\sum_{j=0,1,2}(1+\delta_{jm}(\mathrm{e}^{\mathrm{i}\alpha_3}-1))\ket{j}\bra{j}$ is the qutrit phase gate with respect to the control state $\ket{m}$, and $V \in U(3)$ is a single-qutrit gate that is constrained by the diagonal decomposition of $U$ according to Di and Wei~\cite{Di2013}. Combining (a) and (b), we obtain that the circuit for synthesizing any two-qutrit unitary $U$ consists of only two types of qutrit gates: (i) single-qutrit gates and (ii) qutrit two-level controlled rotation gates.}
    \label{fig-2-qutrit-d3-systhesis}
\end{figure*}

Our ansatz still requires us to encode two-qutrit unitaries into qubit gates. We follow the work~\cite{Di2015} by using single-qutrit gates, controlled diagonal gates and uniformly controlled $R_y$ rotation gates to synthesize a generic $N$-qutrit ($N\geq 2$) circuit. Combined with the method for synthesizing the controlled diagonal gates proposed by Di et al.~\cite{Di2013}, an arbitrary two-qutrit unitary gate $U\in U(9)$ can be synthesized as shown in Figure~\ref{fig-2-qutrit-d3-systhesis}. 

\begin{figure}
    \centering
    \includegraphics[width=\linewidth]{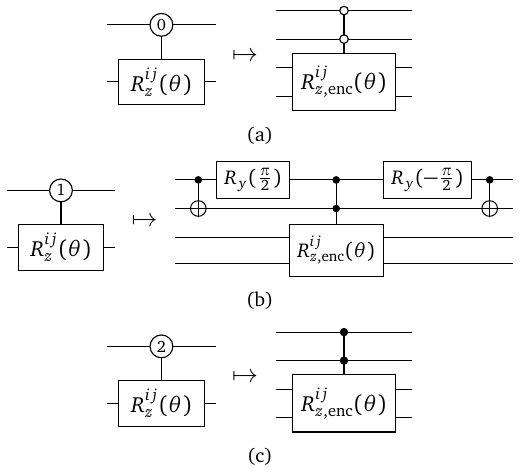}
    \caption{The qubit-based circuit for the encoded qutrit two-level controlled rotation gates in Figure~\ref{fig-2-qutrit-d3-systhesis}. (a) Circuit of encoded qutrit two-level controlled rotation gates when the control state is $\ket{0}$ in Figure~\ref{fig-2-qutrit-d3-systhesis}. (b) Circuit of encoded qutrit two-level controlled rotation gates when the control state is $\ket{1}$ in Figure~\ref{fig-2-qutrit-d3-systhesis}. Here, $R_y$ is the single-qubit rotation gate. (c) Circuit of encoded qutrit two-level controlled rotation gates when the control state is $\ket{2}$ in Figure~\ref{fig-2-qutrit-d3-systhesis}. In all subplots, $R_{z,\text{enc}}^{ij}$ is the two-level rotation gate determined by specializing $V^{ij}_{\text{enc}}$ in Figure~\ref{fig-1-qutrit-encoded-systhesis} with rotation gate $R_z$ for $(i,j) \in \{(0,1),(0,2),(1,2)\}$.}
    \label{fig-qubit-circuit-for-encoded-d3-2-level-CRG} 
\end{figure}

The circuit in Figure~\ref{fig-2-qutrit-d3-systhesis} includes only two types of qutrit gates: (i) single-qutrit gates and (ii) qutrit two-level controlled rotation gates, both of which can be encoded by our gate encoding method. All encoded single-qutrit gates can be decomposed into two-qubit gates using the decomposition in \eqref{decompose 1-qutrit unitary} and synthesis method in Figure~\ref{fig-1-qutrit-encoded-systhesis}. For the encoded qutrit two-level controlled rotation gates, according to the specific state of the control, they can be decomposed into qubit-based gates in Figure~\ref{fig-qubit-circuit-for-encoded-d3-2-level-CRG}. Note that the two-qubit gate $V^{ij}_{\text{enc}}$ in Figure~\ref{fig-qubit-circuit-for-encoded-d3-2-level-CRG} is the two-qubit gate in Figure~\ref{fig-1-qutrit-encoded-systhesis} for $(i,j) \in \{(0,1),(0,2),(1,2)\}$. 

From the state encoding method, we know that the $n$-qutrit Hilbert space is isomorphic to the symmetric subspace of the $2n$-qubit space. From the gate encoding method, we know the encoded qutrit gate is symmetry-preserving. From these plus the universality of qutrit gates proposed by Di et al.~\cite{Di2015}, we conclude that the qubit-based gates in Figure~\ref{fig-1-qutrit-encoded-systhesis} and in Figure~\ref{fig-qubit-circuit-for-encoded-d3-2-level-CRG} form a universal gate set that can approximate arbitrary symmetry-preserving qubit-gates.

\section{Results}
We apply the state and gate encoding methods presented above to our qubit-based PQC to locally simulate qutrit ground states of five types of Hamiltonians in \eqref{Hamiltonian322}. The Hamiltonians have different sets of couplings ${\cal{J}}$ given by five contextuality witnesses shown in Table~\ref{table:5 CW}. Given each contextuality witness, the set of couplings ${\cal{J}}$ in Hamiltonian \eqref{Hamiltonian322} is determined and each of qutrit local observables $\{\sigma_a:a\in\{x,y\}\}$ being a real projective measurement is parameterized as 
\begin{equation}
    \sigma_a(w_a) = (\mathrm{e}^{\sum_{k=1}^{3} w_{ak}S_k})\Lambda_a(\mathrm{e}^{\sum_{k=1}^{3}w_{ak}S_k})^{T},
\end{equation}
where $\Lambda_a$ is a diagonal matrix with entries $\pm 1$, $\{S_k\}$ are the basis of three-dimensional skew-symmetric matrices space, and ${w_a} \equiv (w_{a1}, w_{a2},w_{a3})$ are real parameters. Here, all parameters are denoted as ${W} \equiv (w_x, w_y)$. Then, given a set of parameters $W$, the Hamiltonian is specified and the ground state can be computed using MPS-based algorithms such as  the time-dependent variational principle (TDVP) algorithm \cite{Haegeman2011-TDVP} or the variational uniform matrix product state (VUMPS) algorithm \cite{ZaunerStauber2018-VUMPS-short,Vanderstraeten2019-VUMPS-long}. 

\begin{figure*}
    \centering
    \vspace{-4ex}
    \includegraphics[width=0.98\textwidth]{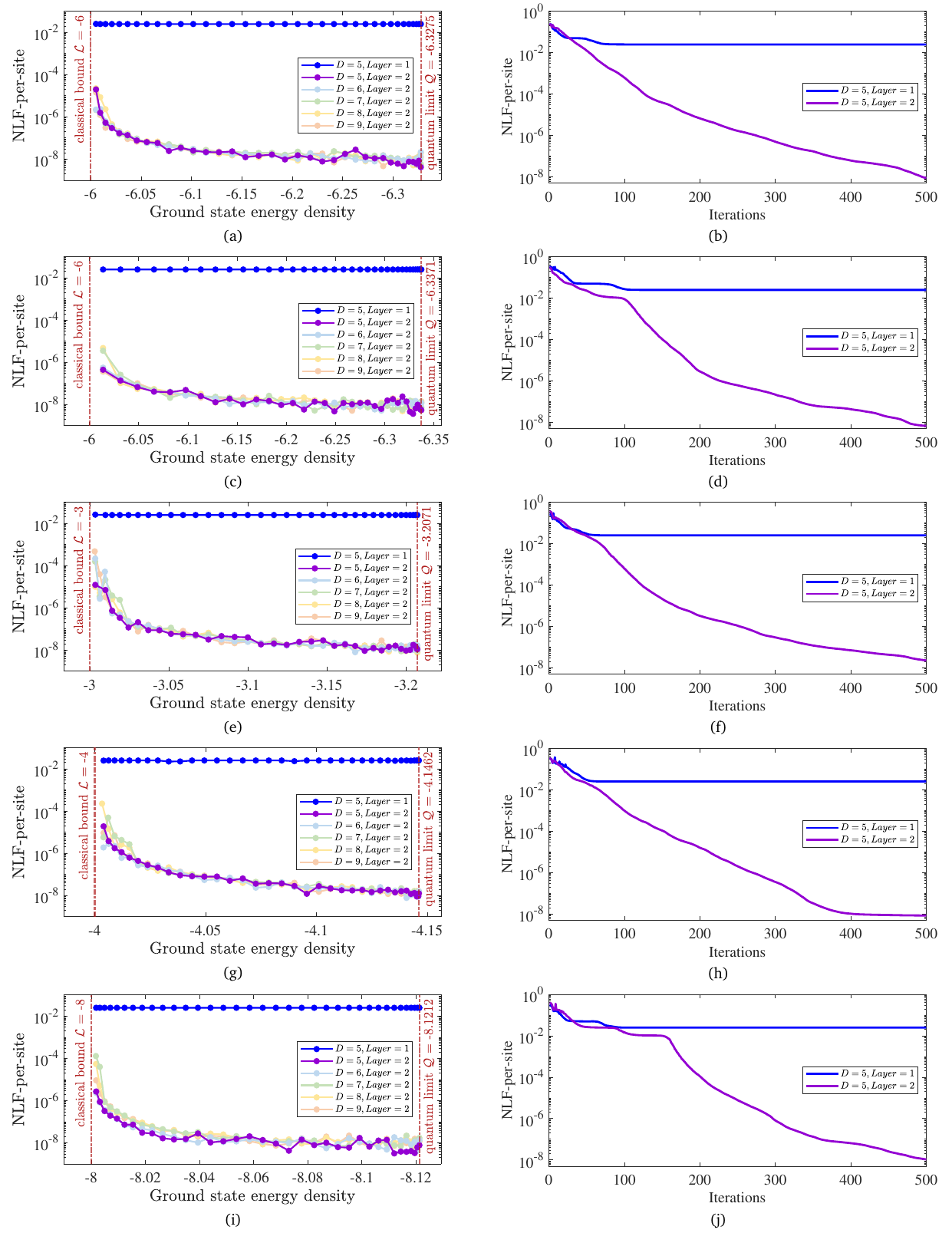}
    \vspace{-3ex}
    \caption{Results for approximating purified 3-site reduced qutrit contextual ground states. These ground states with bond dimensions $D=5,6,7,8,9$ are derived from five contextuality witnesses in Table~\ref{table:5 CW} respectively. (a) No.1 in Table~\ref{table:5 CW}. (b) Points reaching the quantum limit in (a). (c) No.2 in Table~\ref{table:5 CW}. (d) Points reaching the quantum limit in (c). (e) No.3 in Table~\ref{table:5 CW}. (f) Points reaching the quantum limit in (e). (g) No.4 in Table~\ref{table:5 CW}. (h) Points reaching the quantum limit in (g). (i) No.5 in Table~\ref{table:5 CW}. (j) Points reaching the quantum limit in (i).
    All the target states $\ket{\psi_\text{T}}$ are 14-site qubit states and 1-layer and 2-layer quantum circuits in Figure~\ref{fig:7-site-quantum-circuit} are both considered. In subplots (a), (c), (e), (g) and (i), each point represents the ground state of a specific Hamiltonian on the trajectory of the gradient descent aiming at finding the lowest possible ground state energy density, and differences in line color represent different $D$ of the uMPS ground states. Note that points having the same color have precisely the same number of quantum gates and classical variational parameters. In subplots (b), (d), (f), (h) and (j), the convergences of the ground states with $D=5$ achieving the quantum limits are shown.}
    \label{fig:5 results}
\end{figure*}

Beginning with a set of random parameters $W$, the local observables are optimized using gradient descent aiming at finding the lowest possible ground state energy density. We follow the trajectory of the gradient descent and obtain a sequence of Hamiltonians having the same ${\cal{J}}$ but different local observables. From these, we obtain a sequence of qutrit MPS ground states of contextual Hamiltonian exhibiting different amounts of quantum violations. In Figure~\ref{fig:5 results}(a), Figure~\ref{fig:5 results}(c), Figure~\ref{fig:5 results}(e),Figure~\ref{fig:5 results}(g) and Figure~\ref{fig:5 results}(i), each point represents a specific Hamiltonian on the trajectory and different colors represent different bond dimensions $D$ of the uMPS ground states. Due to the constraints of available computational resources, we consider 3-site reduced ground states and five possibilities $D=5,6,7,8,9$ for each model. 
Using the proposed purification procedures, we obtain the purified ground states, which are 7-site qutrit MPSs $\ket{\psi(A,L_1,L_2,R_1,R_2)}$ with respect to $D\in \{5,6,7,8,9\}$. These qutrit states are then generated by the quantum circuits consisting of stacked one layer ($\textit{Layer}=1$) and two layers ($\textit{Layer}=2$)  of two-qutrit unitaries in Figure~\ref{fig:7-site-quantum-circuit} respectively. Here, the number of layers in the quantum circuit dictates part of computational resources for simulating the target quantum state, which are the number of classical variational parameters in classical resources and the number of quantum gates in quantum resources.

To simulate the purified qutrit states on qubit-based quantum circuits, we encode all these 7-site qutrit MPSs $\ket{\psi(A,L_1,L_2,R_1,R_2)}$ into the 14-site qubit states by the symmetric state encoding method, and the encoded qubit states are our target states $\ket{\psi_\text{T}}$. Besides, using the gate encoding method and the circuits for synthesizing symmetry-preserving gates, each two-qutrit gate $U(\boldsymbol{\theta})$ in the quantum circuit in Figure~\ref{fig:7-site-quantum-circuit} can be synthesized by the circuit consisting of qubit gates $U_{\text{enc}}(\boldsymbol{\theta})$ with trainable parameters $\boldsymbol{\theta}$ in Appendix Figure~\ref{fig:two-qutrit-encoded-gate-synthesis}.
Then, fixing the circuit layer of quantum circuit in Figure~\ref{fig:7-site-quantum-circuit}, all the target 14-site qubit states are approximated by the same quantum circuit. In the simulation, each state $\ket{{\psi}_{\text{QC}}}$ generated by the quantum circuit is truncated after 500 iterations. And, we calculate the negative logarithmic fidelity per site (NLF-per-site) ${\cal{F}}$ between the target state $\ket{\psi_\text{T}}$ and $\ket{{\psi}_{\text{QC}}}$:
\begin{equation}
    {\cal{F}} = - \frac{\ln{\sqrt{ \langle{{\psi}}_{\text{QC}} {\ket{\psi_\text{T}}} {\bra{\psi_\text{T}}}{\psi_{\text{QC}}\rangle }}}}{N},
\end{equation}
where $N$ is the number of qubits for the target state. In this situation, if two target states $\ket{\psi^1_{\text{T}}}$ and $\ket{\psi^2_{\text{T}}}$ have the same bond dimension $D$, and $\ket{\psi^1_{\text{QC}}}$ has a lower NLF-per-site than $\ket{\psi^2_{\text{QC}}}$, then we conclude that preparing $\ket{\psi^1_{\text{T}}}$ consumes fewer circuit resources than preparing $\ket{\psi^2_{\text{T}}}$.

We carry out the simulation on the MindQuantum \cite{Xu2024-mindquantum}, which is an open-source, high-performance platform for simulating quantum circuits and algorithms using CPU and GPU acceleration. The simulation results are presented in Figure~\ref{fig:5 results}. In the left side of Figure~\ref{fig:5 results}, i.e., Figure~\ref{fig:5 results}(a), Figure~\ref{fig:5 results}(c), Figure~\ref{fig:5 results}(e), Figure~\ref{fig:5 results}(g) and Figure~\ref{fig:5 results}(i), each figure corresponds to one contextuality witness, and each point in the figure corresponds to one ground state of the Hamiltonian with form \eqref{Hamiltonian322} when the local physical dimension is $3$. Furthermore, each point is the average outcome of optimization runs, calculated from 10 separate repetitions. We compare the results for different circuit layers of quantum circuit in Figure~\ref{fig:7-site-quantum-circuit} and bond dimension $D$ of the uMPS ground states. Note that the minimal bond dimension for the ground states to maximally violate corresponding five many-body witnesses shown in Table~\ref{table:5 CW} is  $D = 5$. We emphasis that all points compared inside one sub-figure, and derived from the circuit having the same circuit layer, have precisely the same number of quantum gates and classical variational parameters. As the plots show that the NLF-per-site decreases as the violation of contextuality witness increases, even though there are minor oscillations caused by some noises from the numerical optimization. This indicates that the stronger contextuality exhibited by the ground state, the less circuit resources are required to prepare the state.

In the right side of Figure~\ref{fig:5 results}, that is Figure~\ref{fig:5 results}(b), Figure~\ref{fig:5 results}(d), Figure~\ref{fig:5 results}(f), Figure~\ref{fig:5 results}(h) and Figure~\ref{fig:5 results}(j), we show the convergence for the ground state reaching the quantum limit of five quantum Hamiltonians in Table~\ref{table:5 CW} respectively. These five figures show that single layer ($\textit{Layer}=1$) in the circuit Figure~\ref{fig:7-site-quantum-circuit} exhibit insufficient expressibility to prepare those optimal quantum states with a reliable fidelity, and at least two layers ($\textit{Layer} \geq 2$) are required. It can be seen that, except for the already converged fourth model, the remaining models could still achieve lower values of NLF-per-site without truncating the number of optimization steps, suggesting the superiority of our method in state preparation with high-precision. Besides, as shown in Table~\ref{table:NLF-per-site classical state}, using the same circuit resources, i.e., $\textit{Layer = 2}$, as generating the contextual ground states, the NLF-per-site of each ground state with the ground state energy density being the classical bound is much smaller than the contextual ones. This indicates that preparing contextual ground states is more expensive than the ground states that are classical. According to Table~\ref{table:NLF-per-site classical state}, we can infer that, under identical circuit structure shown in Figure~\ref{fig:7-site-quantum-circuit}, classical resources and quantum resources, preparing a fully separable state (i.e., a product state) exhibits a lower NLF-per-site compared to other quantum states, indicating that product states are more straightforward to prepare.
\begin{table}
    \caption{The negative logarithmic fidelity per site (NLF-per-site) for noncontextual ground states. Here, each ground state energy density equals to the classical bound meaning that the ground state can not exhibit contextuality. ${\mathcal{L}}$ is classical bound, $D$ is the bond dimension and No. indicates the order of the contextuality witnesses in Table~\ref{table:5 CW} from which the ground states are derived. Each NLF-per-site is the average result of the final converged values from 5 runs of the two-layer circuit optimization.}
    \renewcommand\arraystretch{1.4}
    \setlength{\tabcolsep}{4ex}
    \begin{ruledtabular}
    \begin{tabular}{cccc}
        No. & ${\cal{L}}$ & $D$ & NLF-per-site \\ \hline
        1 & -6 & 1 &  1.850371707708594e-17 \\
        2 & -6 & 1 &  5.551115123125784e-17 \\
        3 & -3 & 1 &  3.700743415417189e-17 \\
        4 & -4 & 1 &  7.401486830834375e-17 \\
        5 & -8 & 3 &  4.434600834700317e-13
    \end{tabular}
    \end{ruledtabular}
    \label{table:NLF-per-site classical state}
\end{table}

\section{Conclusions}
In this paper, we investigate the connection between contextuality and circuit resources required to simulate local approximation to the ground states of maximally contextual quantum Hamiltonians. Since maximal quantum violation demands qutrit systems, we propose a symmetric encoding framework that enables the simulation of any qutrit state using qubit-based parameterized quantum circuits. We find that generating ground states exhibiting contextuality consumes more circuit resources than those that do not exhibit contextuality. We find that, given the same amount of classical and quantum resources, the more contextual a Hamiltonian becomes, the easier it is to faithfully simulate its local ground state. High-dimensional quantum circuit platforms have wide applications in the fields of quantum computing and quantum simulation. Our proposed universal gate set plays a central role in this context. However, considering the synthesis of high-dimensional gates, there is room for improvement in the number of CNOTs, multi-controlled gates, and the corresponding circuit structure in the future.

\section{Data Availability}
The data analyzed in this work are available from the corresponding author upon reasonable request.

\section{Code Availability}
The code used in this work is available from the corresponding author upon reasonable request.

\section{References}
\bibliographystyle{naturemag}
\bibliography{ref.bib}

\section{Acknowledgments}
Z.W. is supported by the Sichuan Provincial Key R\&D Program (2024YFHZ0371), the Shenzhen Institute for Quantum Science and Engineering (SIQSE202106) and the National Natural Science Foundation of China (62250073, 62272259).

\appendix

\section{Supplementary Information for ``Cost of locally approximating high-dimensional ground states of contextual quantum models''}

\begin{figure*}
    \centering
    \includegraphics[width=\linewidth]{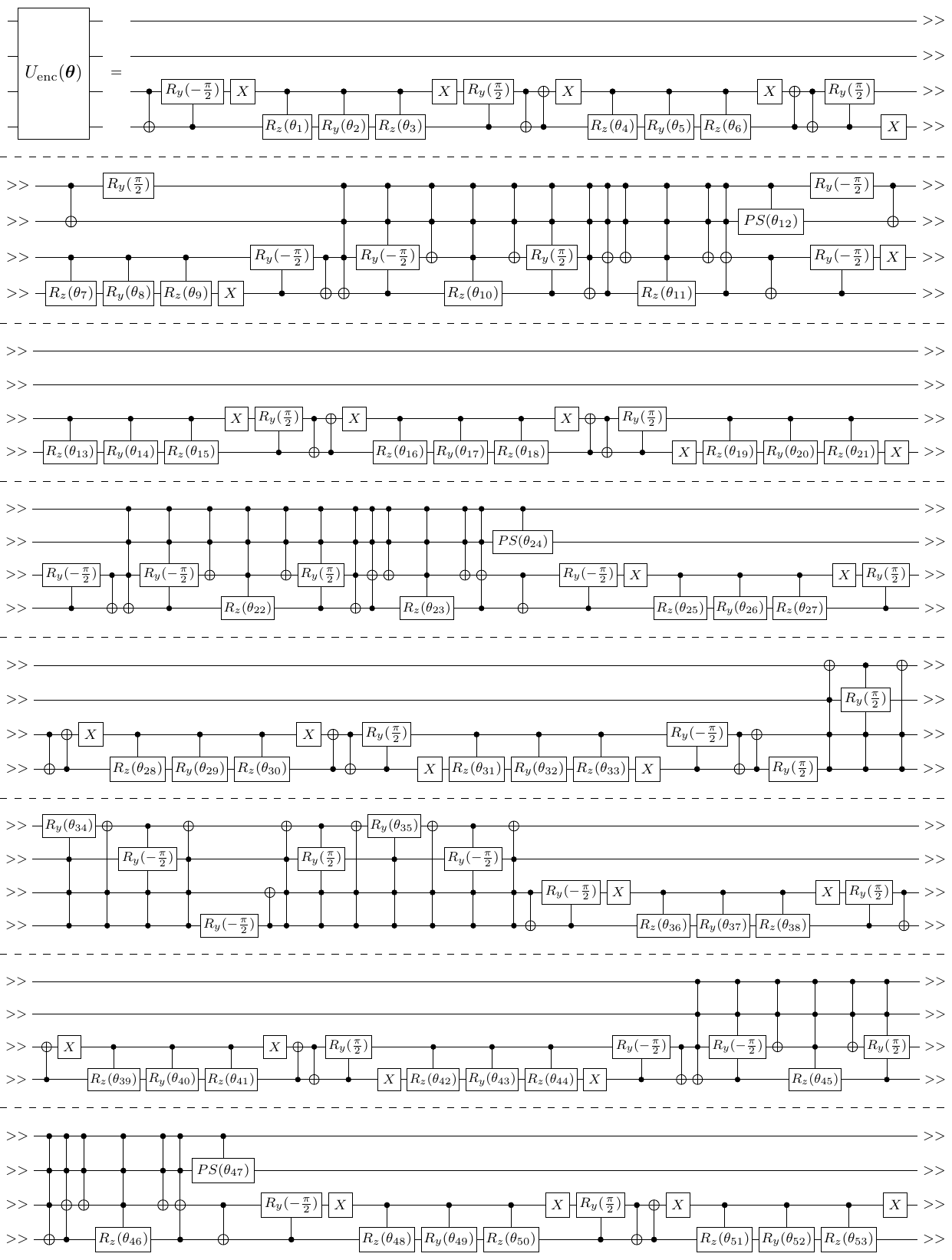}
\end{figure*}
\begin{figure*}
    \centering
    \includegraphics[width=\linewidth]{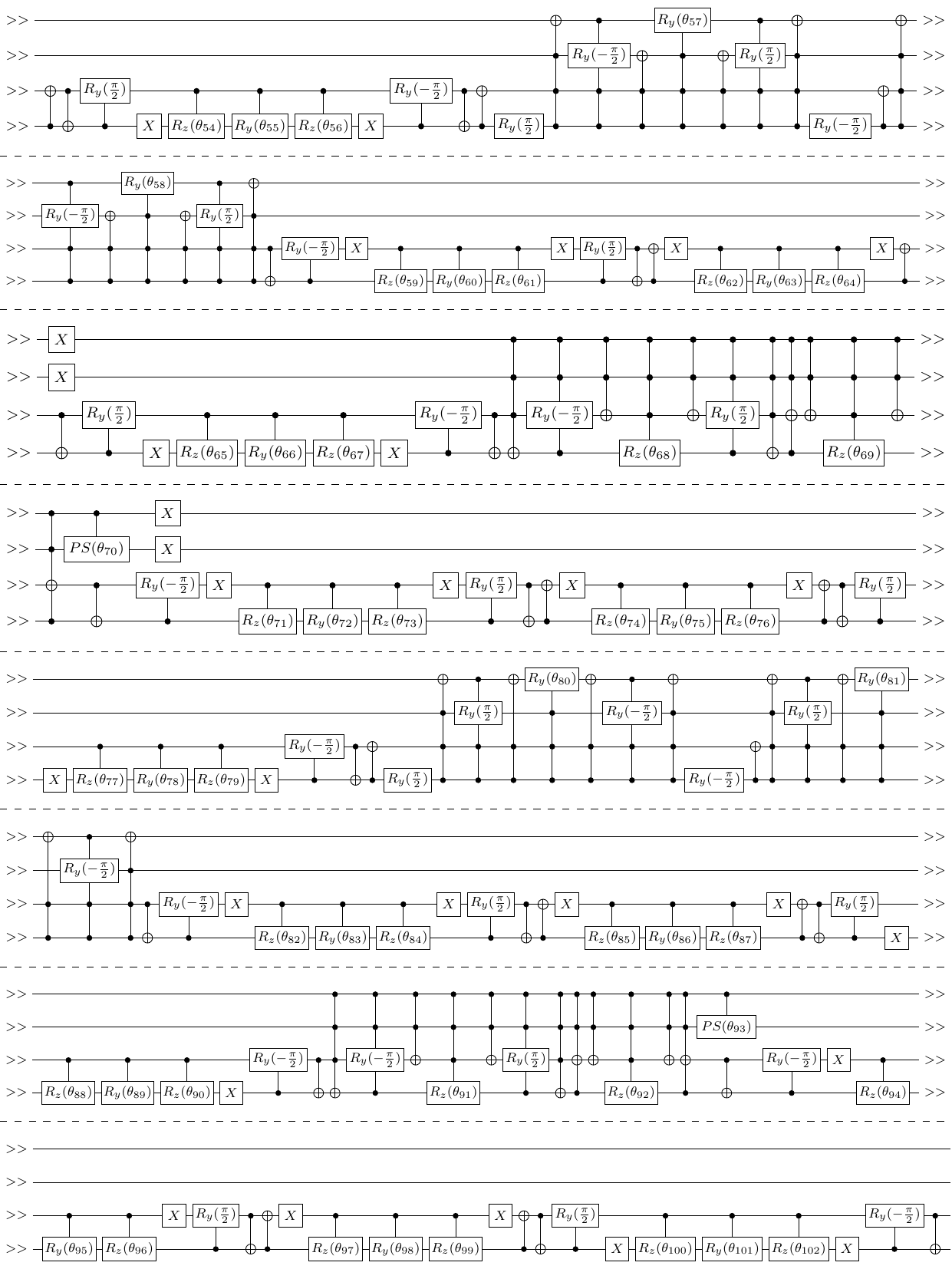}
    \vspace{-3ex}
    \caption{Synthesis of symmetry-preserving four-qubit gate. An illustration of the symmetry-preserving four-qubit circuit for synthesizing an arbitrary two-qutrit unitary gate $U({\boldsymbol{\theta}}_i^k)$ in Fig.~1 for $i\in \{1,2,\dots,6\}$ and $k\in\{1,2,\dots,K\}$, where $PS(\theta)$ is phase shift gate.}
    \label{fig:two-qutrit-encoded-gate-synthesis}
\end{figure*}

\end{document}